\documentclass[aps,prl,showpacs]{revtex4}

\usepackage{amsmath,epsfig,amsfonts,epsfig,graphicx}

\begin{document}

\title{Community Structure in Congressional Cosponsorship Networks}
\author{Yan Zhang$^{1}$, A. J. Friend$^{2}$, Amanda L. Traud$^3$, Mason A. Porter$^{4}$
,\\James H. Fowler$^{5}$, and Peter J. Mucha$^{3,6}$ \\
  $^1$\footnotesize{Department of Mathematics, California Institute of Technology, Pasadena, CA 91125, USA}\\  
  $^2$\footnotesize{School of Mathematics, Georgia Institute of Technology, Atlanta, GA 30331-0160, USA}\\
  $^3$\footnotesize{Department of Mathematics, University of North Carolina, Chapel Hill, NC 27599-3250, USA} \\ 
  $^4$\footnotesize{Oxford Centre for Industrial and Applied Mathematics, Mathematical Institute, University of Oxford, OX1 3LB, UK} \\
   $^5$\footnotesize{Department of Political Science, University of California at San Diego, La Jolla, CA 92093-0521, USA} \\
  $^6$\footnotesize{Institute for Advanced Materials, University of North Carolina, Chapel Hill, NC 27599-3250, USA}
  }

\begin{abstract}
We study the United States Congress by constructing networks between Members of Congress based on the legislation that they cosponsor. Using the concept of modularity, we identify the community structure of Congressmen, as connected via sponsorship/cosponsorship of the same legislation, 
to investigate the collaborative communities of legislators in both chambers of Congress.  This analysis yields an explicit and conceptually clear measure of political polarization, demonstrating a sharp increase in partisan polarization which preceded and then culminated in the 104th Congress (1995-1996), when Republicans took control of both chambers.  Although polarization has since waned in the U.S. Senate, it remains at historically high levels in the House of Representatives.
\end{abstract}


\maketitle



\section{Introduction}


Party politicians in the United States have become more polarized over the last 20 years, which is leading in turn to a gradual polarization of the electorate \cite{pr01,hether,saunders}.  However, voters are not as polarized as portrayed by the media \cite{dimag}.  Thus, although the 1994 Congressional elections saw a ``Republican Revolution'' that ended forty years of Democratic majorities in the House of Representatives (the longest span of single-party rule in Congressional history \cite{revolution}), it has been argued using the analysis of roll call votes that this change reflected a gradual polarization in U.~S. politics \cite{pr01}.  These arguments are based on a simple \emph{ad hoc} measure of polarization that is simply the mean difference in ideological locations of members of the Democratic and Republican parties \cite{mccarty2006}.

In this paper, we study Congress using a different set of tools---those of network theory, which have been successfully employed to characterize a wide variety of complex systems \cite{str01,newmansirev}.  Recent work has illustrated potential insights from analyzing Congress as a social network: Members of Congress who are more ``central" tend also to be more important politically \cite{fowlershort,fowler} and ``communities"  of committees and subcommittees can be identified without specific political knowledge about the committees or their members \cite{congshort,conglong,conggallery}.  We show here that investigating the organizational structure of Congress using the idea of ``modularity'' \cite{structeval,newmod,newmodlong,resolution} is particularly effective at identifying and analyzing the historical development of communities of legislators.  In particular, it can be used to study partisan polarization in Congress directly from the network data without the need to supply specific information about the ideology or political orientations of the legislators themselves, the committees on which they sit, or the legislation they support.

 \section{Legislation Cosponsorship Networks}

In the U.S. Congress, legislators can make public their support for a particular bill by cosponsoring it.  The act of cosponsorship is simple---a legislator simply signs his or her name to a bill that has been introduced for consideration in the chamber.  This has caused some political scientists to disregard the act of cosponsorship as ``cheap talk'' \cite{wilson1997}.  However, the average legislator cosponsors only 2--3\% of all possible bills \cite{fowler}, so the tough part is deciding \emph{which} bills merit support.  Legislators themselves clearly think the act is important, because they expend considerable effort recruiting cosponsors with personal contacts and ``Dear Colleague'' letters, and they frequently refer to cosponsorships in floor debate, public discussion, letters to constituents, and campaigns \cite{Campbell1982}.

Our primary interests are the Congressional networks defined by legislation cosponsorship in the U.S. Senate and House of Representatives from the 96th--108th Congresses (1979--2004), a time frame during which the cosponsorship rules remained relatively unchanged in each legislative body.
We define legislation to encompass all resolutions, public and private bills, and amendments, and we treat sponsorship and cosponsorship on equal footing for simplicity.  We investigate each two-year term of Congress separately, yielding thirteen separate cosponsorship networks for each chamber of Congress.  In these two-mode (``bipartite") networks, a Member of Congress is connected by an edge to each bill he/she sponsored or cosponsored.  This is encoded using a bipartite adjacency matrix $\mathbf{M}$, with entries $M_{ij}$ equal to $1$ if legislator $i$ (co-)sponsored bill $j$ and $0$ if not.  That is, the two types of nodes are Congressmen and bills, and each edge in the network represents a sponsorship or cosponsorship.

Another important feature of legislative organization is the structure of committee and subcommittee assignments.  Before legislation is considered on the floor of the chamber, it is usually assigned to committees that have jurisdiction based on the issues the bill addresses.  \emph{Standing} committees are permanently established by the rules of each chamber, whereas \emph{select} committees are established by resolutions and might not be permanent.  The partisan balance in each committee (i.e., the numbers of Democrats and Republicans) typically reflects the partisan balance of the whole chamber, and each party controls which of its legislators are nominated for which committees.  Once committees are established, they may divide themselves into subcommittees with narrower jurisdictions.

We can use information about these Congressional committee and subcommittee assignments to create another kind of network (again considering each two-year term separately).  For this collection of networks, a unit value of the entry $\tilde{M}_{ij}$ of a bipartite adjacency matrix indicates the assignment of Representative $i$ to committee or subcommittee $j$.  We treat parent committees (including both standing and select committees) and subcommittees without distinction.  



\begin{figure}
\begin{center}
\includegraphics[width=16cm]{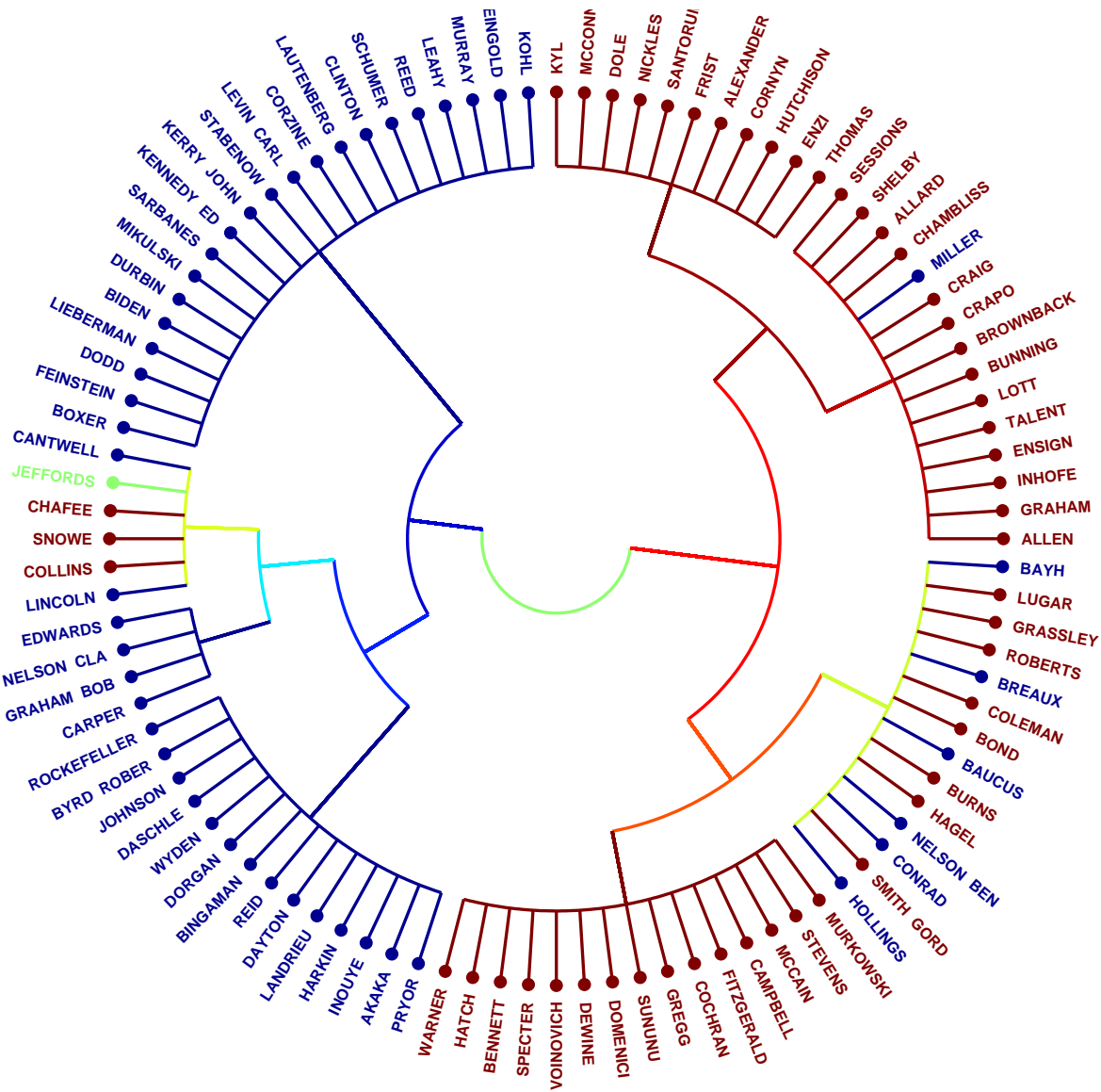} 
\caption{[Color] Community structure in the legislation cosponsorship network of the 108th Senate (2003-2004), as seen using a tree (or ``dendrogram") produced from the original network.  This dendrogram reveals a strong polarization between Republicans (red) and Democrats (blue).  (James Jeffords, an Independent, is shown in green.)
}
\label{legislatesenate}
\end{center}
\end{figure}

We analyze the cosponsorship networks using one-mode (``unipartite") projections with adjacency matrices $A_{ij} = \sum_k M_{ik}M^T_{kj}$, in which nodes are legislators and the weighted edges connecting them indicate how many bills they together (co-)sponsored (in the committee assignment networks, a weighted edge indicates the number of committees and subcommittees on which two legislators both sit).  To identify network communities \cite{commreview,structpnas,structeval,vicsek}, we use the intuitive fact, embodied by {\it modularity} \cite{structeval}, that a community should have more internal connections among its nodes than connections between its nodes and those in other communities.  Specifically, the modularity $Q$ is defined as the fraction of the edge weight contained within the specified communities minus the expected total weight of such edges (under standard, suitable assumptions \cite{newmod,newmodlong,np}).  That is,
\begin{equation}
	Q = \frac{1}{2m}\sum_{ij}\left[ A_{ij} - \frac{k_ik_j}{2m}\right]
	\delta(g_i,g_j)\,,
\end{equation}
where $m$ is the total weight of the edges in the network, $k_i$ is the (weighted) degree of the $i$th node, $g_i$ is the community to which $i$ belongs, and $\delta(g_i,g_j)=1$ if $i$ and $j$ belong to the same community and 0 otherwise.  Modularity, computed for selected partitions of the network, thereby measures the extent to which the identified interactions between legislators take place within the identified community partitions rather than across them.  We employ a slight modification \footnote{We recursively subdivide the network using the leading-eigenvector method (without any refinement) described in Ref.~\cite{newmodlong} until the modularity of each of the obtained subnetworks cannot be increased by further partitioning via leading eigenvectors.
We then obtain additional partitioning by treating each existing partition as if it had no external connections (i.e., as if it were itself the full network of interest).} of the leading-eigenvector community-detection method presented in Ref.~\cite{newmodlong}, recursively partitioning each network to generate trees or ``dendrograms" that convey the hierarchical structure of the network.
This process identifies communities of various sizes via the clusters of legislators formed at each stage of the iterative partitioning algorithm.

Figure \ref{legislatesenate} depicts the dendrogram for the cosponsorship network of the 108th Senate.  As this figure illustrates, the initial partitioning of the cosponsorship networks into two communities yields one group consisting predominantly of Republicans and another consisting predominantly of Democrats.  We find that this is the case in each of the cosponsorship data sets for both the Senate and the House of Representatives. 
However, this partitioning does not lie precisely along party lines.  Our analysis picks out known moderate Senators who collaborate more with members of the opposite party, confirming recognized political behavior without incorporating any specific knowledge about their political orientations. For example, Fig.~\ref{legislatesenate} indicates that several liberal Republicans, such as Lincoln Chafee [R-RI], Olympia Snowe [R-ME],
and (former Republican) James Jeffords [I-VT], appear to be grouped with the Democrats; whereas several conservative Democrats, such as Zell Miller [D-GA], John Breaux [D-LA], and Kent Conrad [D-ND], appear to be closely connected to the Republicans.  It is well known that politicians like these frequently vote with the opposite party \cite{pr,voteview}, but our analysis shows that this partisan mixing actually occurs much earlier (i.e., when they collaborate on cosponsoring legislation).


\section{Network Modularity and Partisan Polarization}

\begin{figure}
\includegraphics[width=8.0cm]{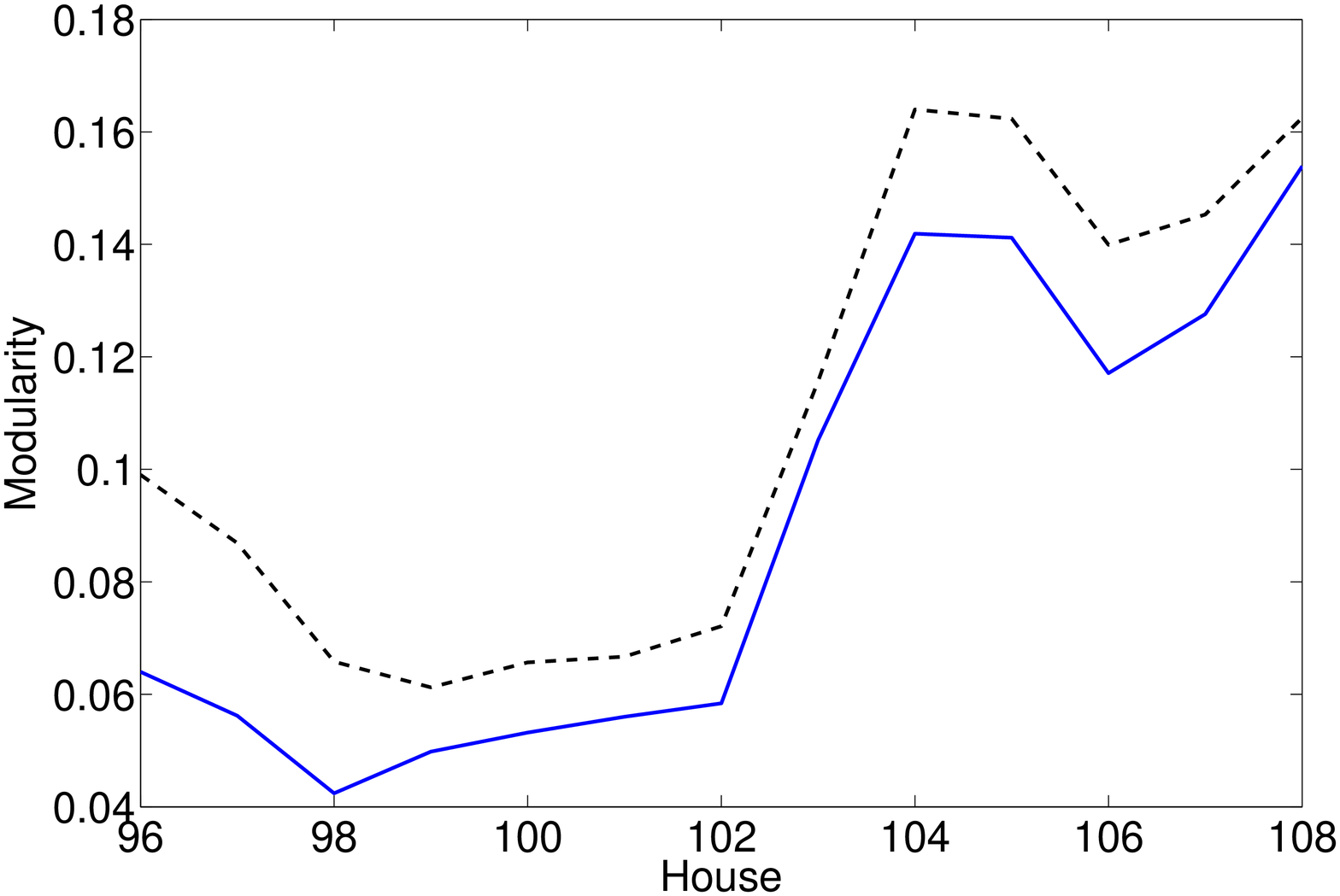} 
\includegraphics[width=8.0cm]{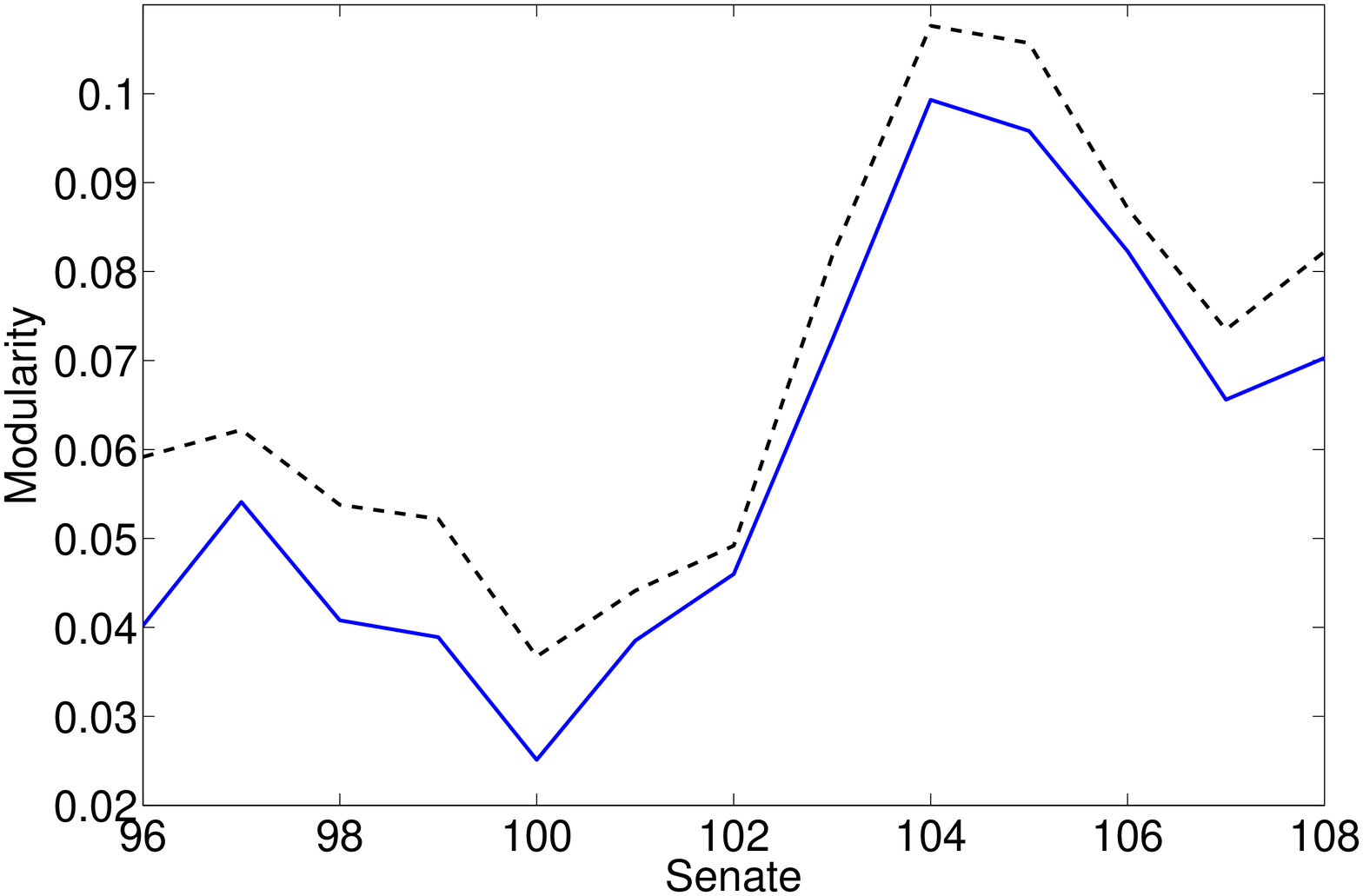} 
\caption{[Color online] Increase in party polarization from the 96th term of Congress (1979-1980) to the 108th term (2003-2004), as seen using modularity measures.  The panels show the party modularity (solid curve) and maximum modularity obtained by leading eigenvectors (dashed curve) for the 96th--108th House of Representatives (Left) and Senate (Right).  In both the House and the Senate, the party modularity approaches the maximum modularity in more recent Congresses, indicating that the natural split of the legislation cosponsorship network is aligning more closely with party affiliation.
}
\label{mods}
\end{figure}

Because modularity measures the number of intra-community versus inter-community edges for a given partition, it can 
be used to quantify the increase in polarization in the U.S. Congress.  This is an important conceptual improvement over existing measures that simply report the mean difference in ideology between the two major parties \cite{mccarty2006}, because modularity does not depend on the assumption that the parties themselves are the relevant communities.  In Fig.~\ref{mods}, we plot for both the House (left panel) and Senate (right panel) the modularity obtained for the first network split into two partitions (dashed curves), which gives the maximum modularity of any partitioning into communities as obtained by the leading-eigenvector method for each of our 26 data sets. Strictly speaking, there may be partitions with even larger modularity (finding a global maximum for modularity is known to be an NP-complete problem \cite{np}), but for simplicity we will hereafter use the term ``maximum modularity'' to indicate the largest value obtained by the leading-eigenvector method. We also plot the modularity obtained by partitioning the network according to political party (solid curves).  By convention, we place all non-Democrats with the Republicans; other placements of Independents have only slight effects on the modularity values.  As shown in the figure, the modularity is relatively steady at first, rises sharply at the 103rd Congress, peaks at the 104th--105th, and then slowly decreases in the Senate while leveling off (or even continuing to increase a little) in the House.  The relatively large modularity obtained by partitioning along party lines (as compared to the maximum modularity) indicates that in some cases simple party identification yields almost as good a partition as the leading-eigenvector method.

Our modularity computations show clearly that partisan polarization in Congress has increased during the past 20 years, with sharp increases both immediately prior to and following the election for the 104th Congress, in which Republicans took control of both the House and Senate.  This suggests that polarization may have been partially a cause---rather than merely an effect---of the partisan change in Congress.  Note additionally that polarization in the Senate has declined from its peak, whereas it has remained near its peak in the House.  This result is consistent with arguments that innovations in House redistricting are a force behind increasing polarization \cite{coxkatz}, as Senate districts conform to unchanging state boundaries whereas House districts are redrawn every 10 years. Finally, as shown in Fig.~\ref{mods}, partisan polarization appears to be contributing to an increasingly large share of total polarization, especially in the House.  One can see this in the figure by observing that the modularity curves become much closer together as a function of time.  This can be quantified using the ratio of party modularity to maximum modularity, which increases from $.0640/.0990 \approx .6465$ in the 96th House to $.1539/.1625 \approx .9471$ in the 108th House and from $.0402/.0592 \approx .6791$ in the 96th Senate to $.0703/ .0822 \approx .8552$ in the 108th Senate.  For both the House and the Senate, the party and maximum modularities each peak during the 104th Congress; the ratios of party to maximum modularity in this term are $.1419/.1640 \approx .8652$ and $.0993/.1076 \approx .9229$. 

Dendrograms produced using modularity eigenvectors reveal larger communities in the legislation cosponsorship networks than in the committee assignment ones, reflecting in part the known high \emph{dimensionality} of the former \cite{talbert}.  This dimensionality reflects the fact that there are many different kinds of overlapping coalitions that can form on many different kinds of issues.  Multidimensional scaling techniques like NOMINATE \cite{pr01} and singular value decomposition (SVD) show that a matrix of roll call votes can be approximated well with a generic liberal-conservative dimension and a second ``social'' dimension \cite{voteview,congshort,conglong}, but the same techniques show that several dimensions are needed to adequately approximate a matrix of cosponsorships.  Political scientists have speculated that the reason for this is that the initial issue space is huge---any legislator can sponsor or cosponsor any policy idea---but the parties take greater control as a bill nears passage, so that recorded votes are strongly influenced by the primary liberal-conservative dimension that spans the two parties \cite{talbert}.

\begin{figure}
\includegraphics[width=8.0cm]{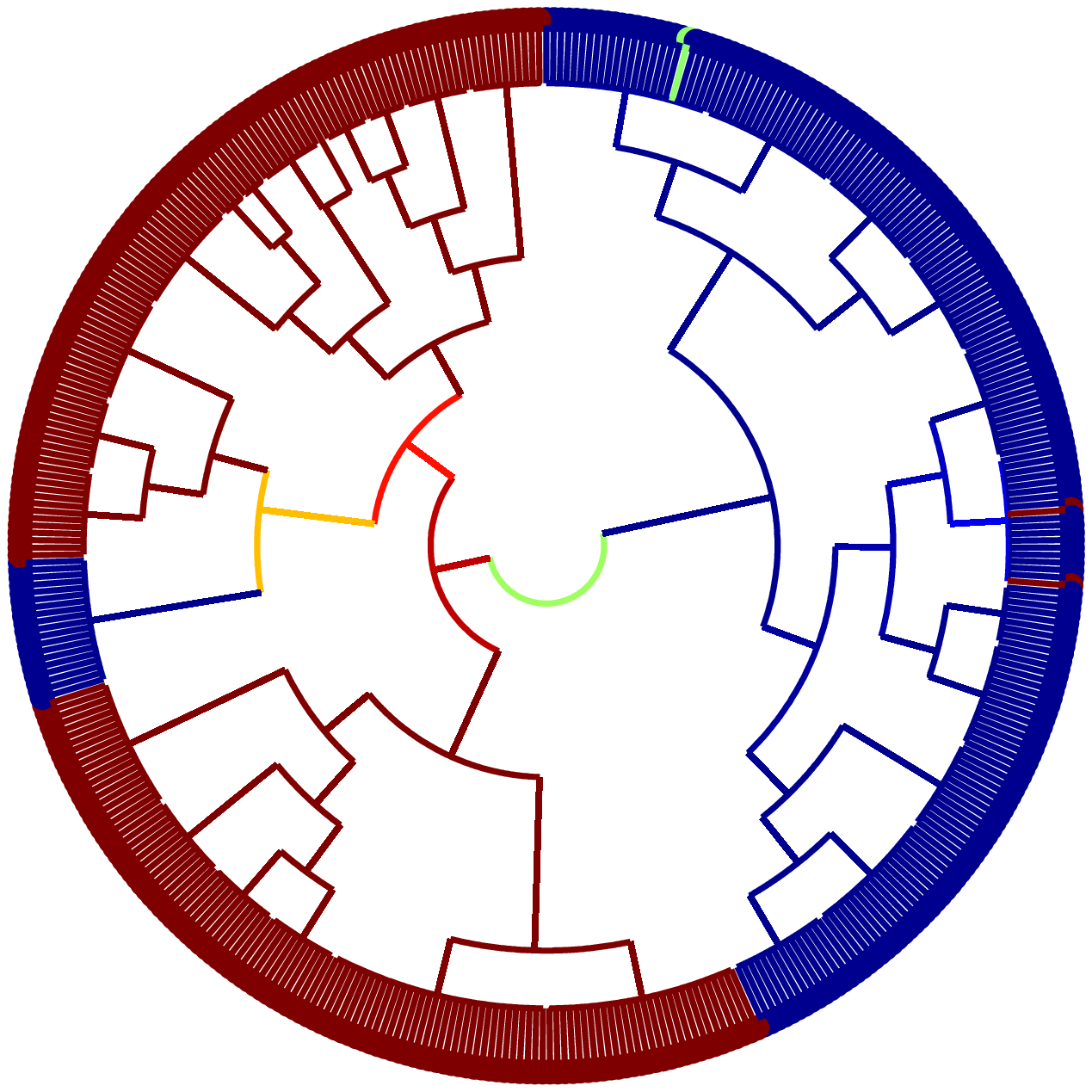} 
\includegraphics[width=8.0cm]{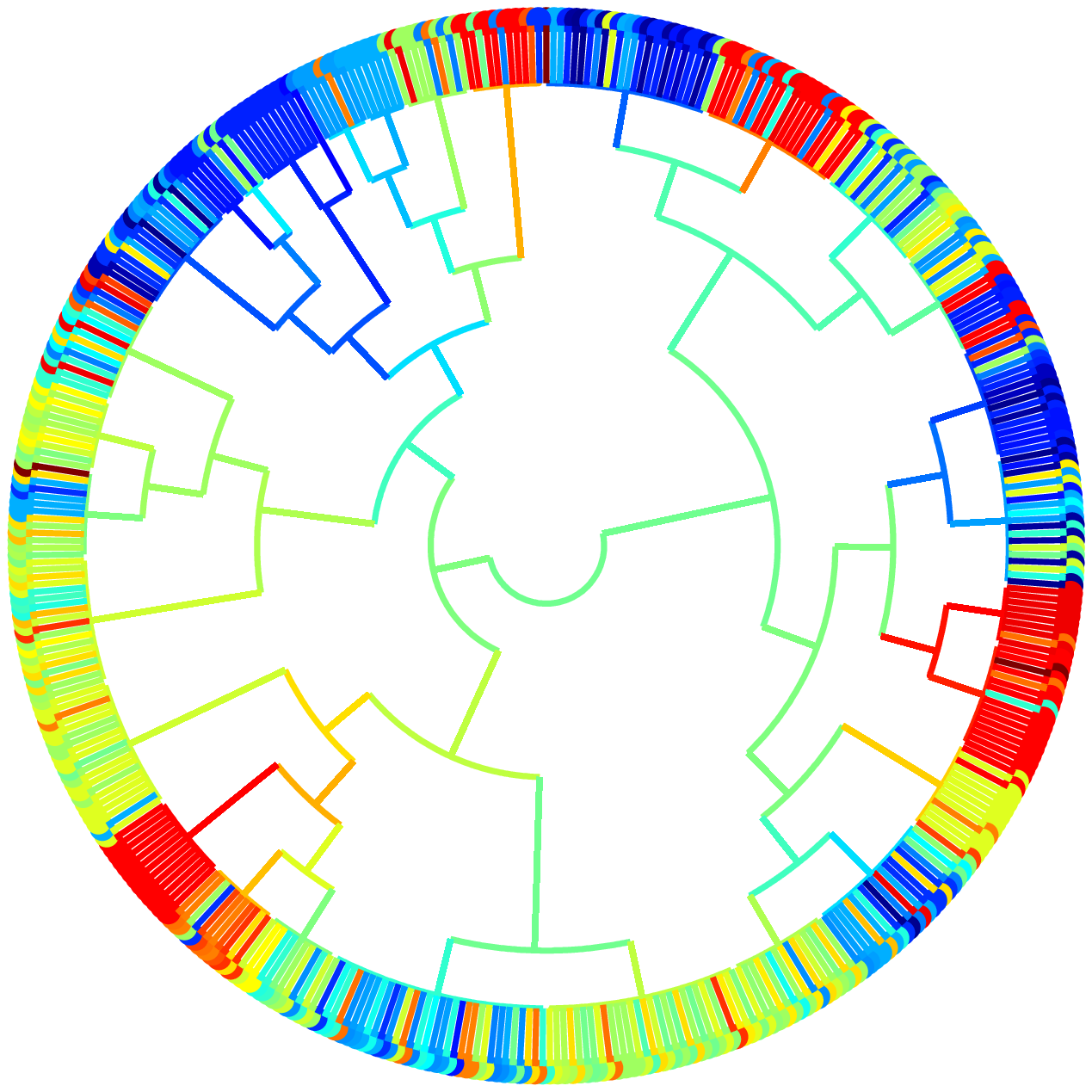} 
\includegraphics[width=8.0cm]{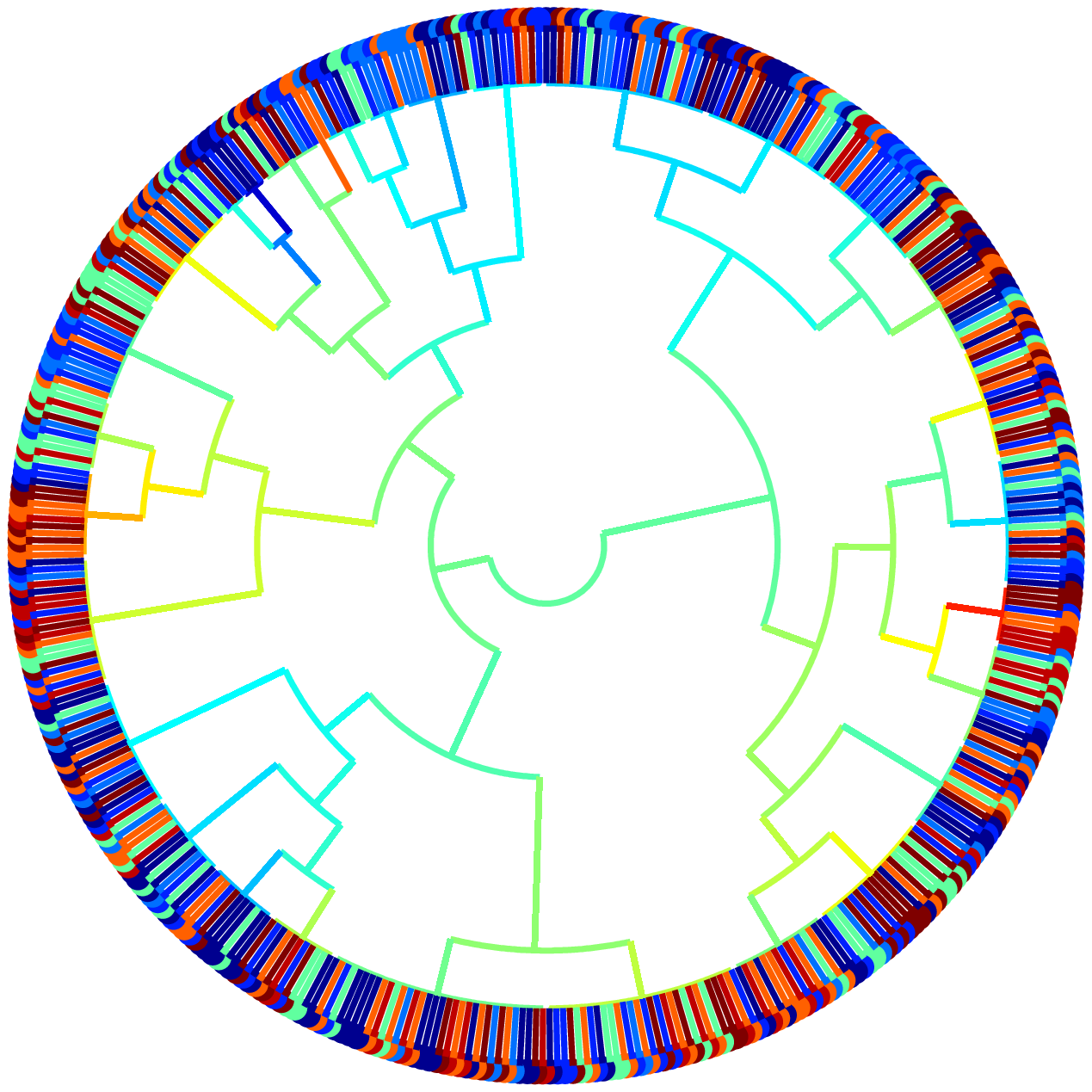} 
\includegraphics[width=8.0cm]{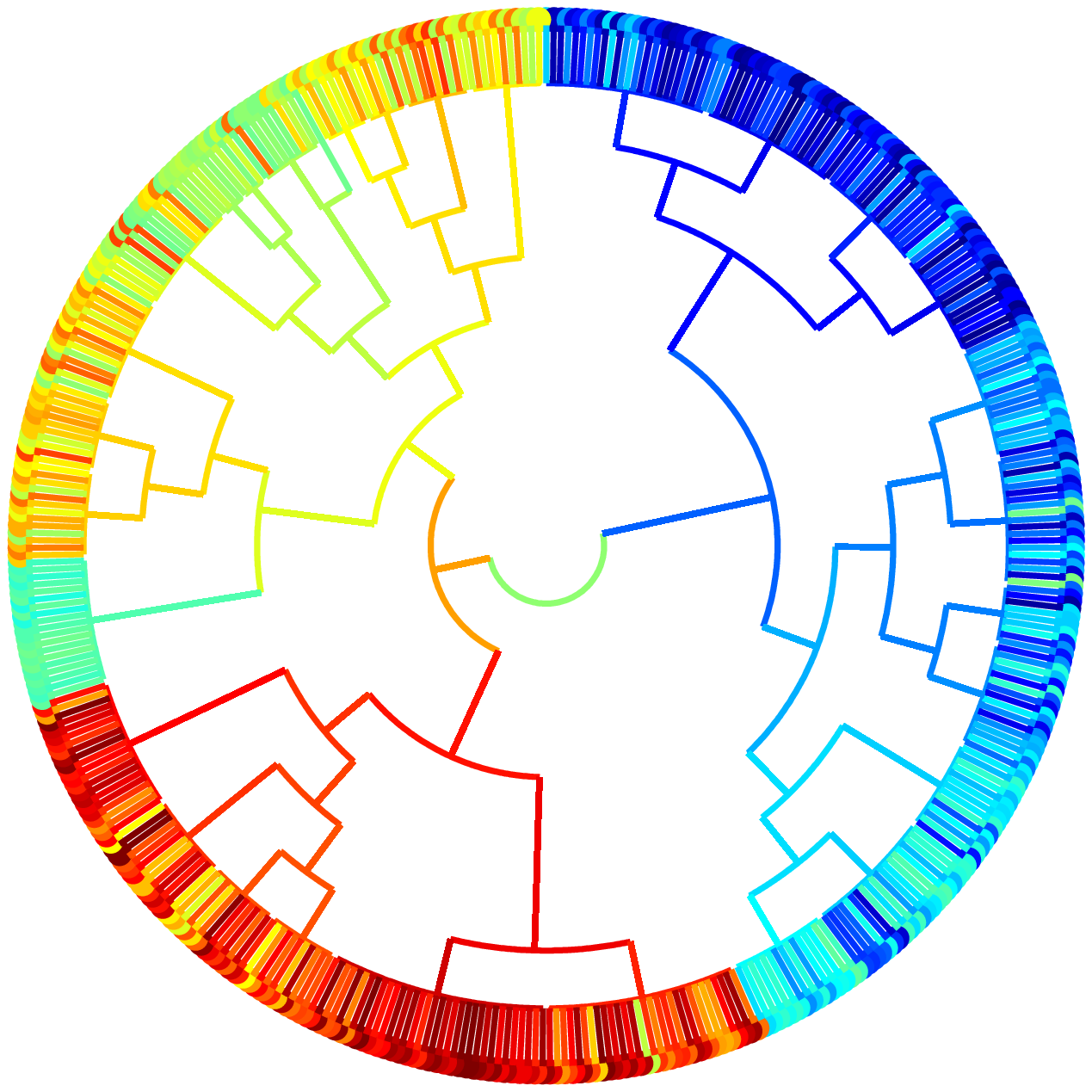} 
\caption{[Color] Dendrogram of the 108th House cosponsorship network, with Representatives colored by political party (upper left panel), state (upper right), community in the committee assignment network (lower left), and roll call voting (lower right).  The correlations indicated by these figures are similar to those in the Houses and Senates in other Congressional terms.  The upper left dendrogram has Democrats in blue, Republicans in red, and Independents in green.  The colors in the upper right dendrogram range from blue to red according to Inter-University Consortium for Political and Social Research (ICPSR) state codes.  (Hence, many geographically close states are shown in similar colors.)  The dendrogram in the lower left is colored by committee community membership as determined by applying the same leading-eigenvector method to the committee assignment network.  The dendrogram in the lower right is colored by the DW-NOMINATE rank-ordering \cite{voteview} computed from House roll call votes.
}
\label{legislatehouse}
\end{figure}

The large communities detected by the leading-eigenvector method correspond to known political cliques (see Fig.~\ref{legislatehouse}, which shows dendrograms for the 108th House).  The strongest correlation is with party (upper left panel), but the upper right panel reveals that the computed communities also have a positive correlation to state, as most communities encompass nodes (Congressmen) with similar or even identical colors (nearby states are colored similarly).  This is reasonable, as many of the bills and amendments involve geography-specific ``pork'' that benefits the (co-)sponsors' districts and regions.  Furthermore, examining the state and party dendrograms together reveals a group of Southern Democrats (from the former Confederate states plus Kentucky, Oklahoma, and West Virginia) that consistently cosponsor with Republicans.  This group, which appears near the 9 o'clock position in the panels of Fig.~\ref{legislatehouse}, starts as a very large bloc in the earlier Congresses studied here 
but decreases to a much smaller group by the later ones, indicating again the gradual increase of partisan polarization \footnote{The Southern Democrat bloc had 69 people in the 96th House; its size then stayed above the mid 50s for several Congresses (reaching a high of 74 Representatives in the 98th House); it contained 43 people in the 102nd House and has been at 20 or less in every House since the 103rd.}.  

In the lower left panel of Fig.~\ref{legislatehouse}, we color the House legislation communities via community membership determined at the maximum-modularity partition obtained by applying the leading-eigenvector method to the Representatives in the House committee assignment network.  Observe at the final splits in the dendrogram that some small clusters of Congressmen from the same committee assignment communities are also grouped together in the cosponsorship network.  Such correspondences seemingly reflect the fact that legislators interact more with fellow members of their own committees and subcommittees (the people with whom they have more contact).  Several contrasting theories of committee assignment have been developed in the political science
literature (see, for example, \cite{nisk,gill,kreh,cox,shep,org}), but these have focused almost exclusively on the institutional explanations, like the role of the party or of specific rules adopted by the chambers to regulate the assignment process.  However, this literature has largely ignored the effects of personal influence and social relationships on the assignment process and vice versa.  In previous work by some of the present authors, we have used community-detection techniques to show that influential Congressional committees appear to be ``stacked'' with partisan party members \cite{congshort,conglong}.  The correlations with legislation cosponsorship reflected in Fig.~\ref{legislatehouse} suggest that this can have a profound impact on the drafting of bills.  

In the lower right panel of Fig.~\ref{legislatehouse}, we compare our observations directly to prior analyses of roll call votes by coloring the legislation cosponsorship dendrograms according to DW-NOMINATE rank-ordering \cite{pr,voteview}.  DW-NOMINATE is one of the modern incarnations of NOMINATE.  As briefly mentioned earlier, NOMINATE is a multi-dimensional scaling technique that is based on iterative algorithms that search for the best-fitting legislator ideologies and bill ideologies that would explain the observed set of votes.  It produces results very similar to those generated by an SVD and it has become the standard technique used in the political science community to measure political ideology from roll call data \cite{pr01,mccarty2006}.  
As expected, Members of Congress with similar ideologies cosponsor a lot of the same legislation and are grouped together in the identified communities.  In this plot, the Southern Democrats (just below the 9 o'clock position) 
are grouped near members of the opposite party and shown as moderates.
Their DW-NOMINATE scores tend to lie close to the median, as moderate Representatives vote with their own party on party-line legislation but vote against their party on many other issues. 


\begin{figure}
\begin{center}
\includegraphics[width=16cm]{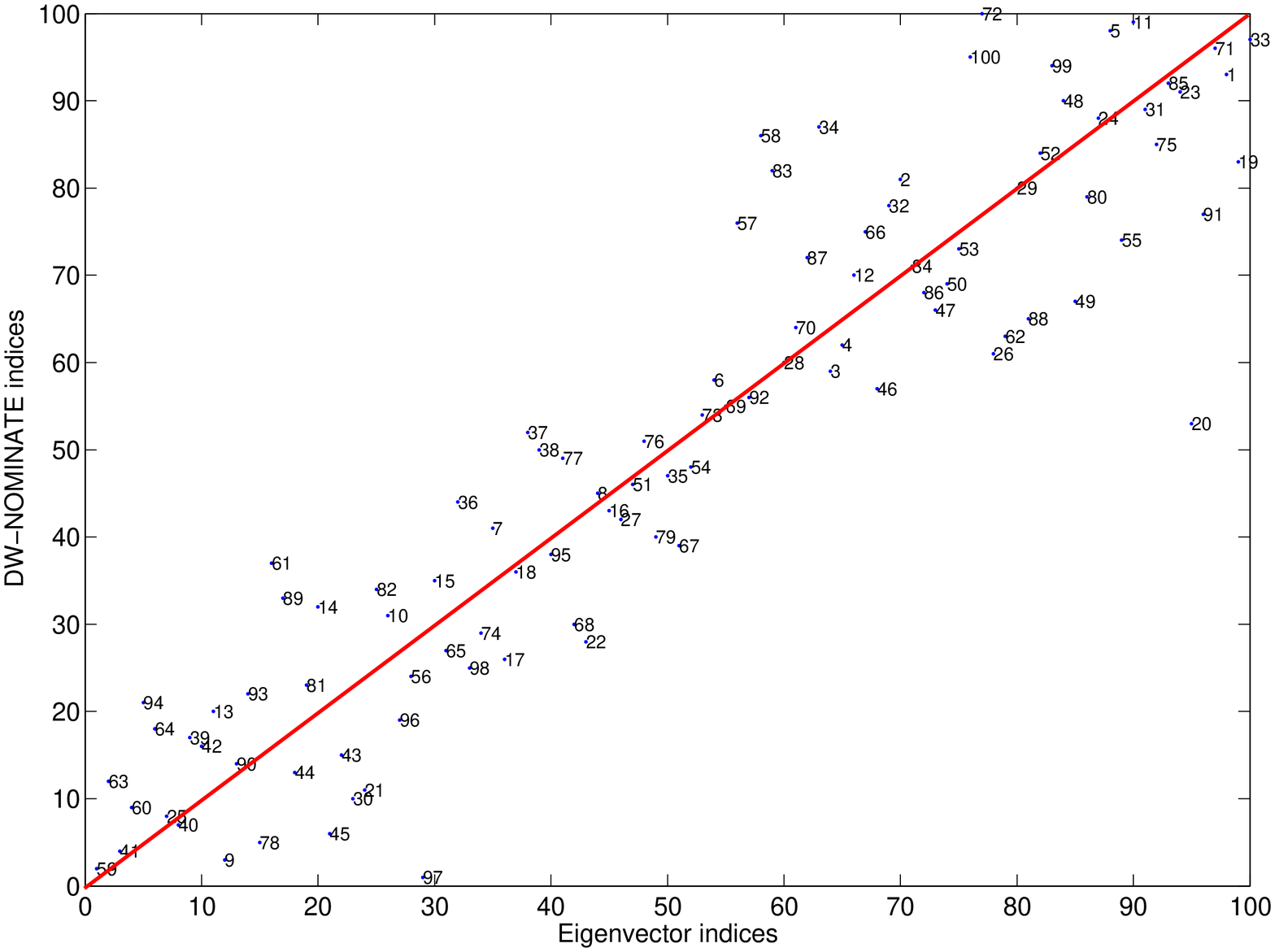} 
\caption{[Color online] Scatter plot of leading modularity eigenvector rank order versus DW-NOMINATE rank order for the 108th Senate.}
\label{scatter}
\end{center}
\end{figure}

The concept of network modularity provides a fresh perspective on investigating collaborative groups in Congress.  Modularity can also be used in direct quantitative analyses of the legislation cosponsorship networks, providing complementary insights to more traditional multidimensional scaling techniques.  For example, the elements of the leading modularity eigenvector 
allow us to construct a rank-ordering of legislators based on cosponsorship patterns, which we compare with the rank-ordering obtained by DW-NOMINATE on Congressional roll call votes \cite{voteview}.  As we illustrate using a scatter plot for the 108th Senate (see Fig.~\ref{scatter}), these methods give highly correlated rank-orderings (with $R^2$ values typically higher than 0.8) even though they are constructed using different data sets.  The two rank-orderings include roughly half of the same legislators among their Leftmost and Rightmost 10\%, with an average of 23.54 (of 44) matches on the Left (with a standard deviation $\sigma \approx 3.58$) and 21.69 on the Right ($\sigma \approx 3.22$) for the 96th--108th Houses and 5.69 (of 10) on the Left ($\sigma \approx 1.44$) and 5.46 on the Right ($\sigma \approx 1.40$) for the 96th--108th Senates.  

For every term of Congress, we also define an absolute rank difference for each legislator (with multiple entries for legislators who held office during more than one term) between his or her eigenvector and DW-NOMINATE rank orderings.  For both the Senate and the House (calculated separately), we compute an average difference by adding the absolute rank differences of all Congressmen and dividing by the total number of Senators or Representatives (counting multiplicities for Congressmen who held office during more than one term).  We find that the House rankings produced using the eigenvector method differ on average by about 39.96 Representatives (about 9.15\%) from those produced by DW-NOMINATE and that the Senate rankings differ on average by about 8.29 Senators (about 8.20\%, recalling that some terms have more than 100 Senators because of mid-term replacements). These results validate the use of this network-modularity method and suggest that it is possible to derive ideology measures from cosponsorship data in spite of its known high dimensionality \cite{talbert}.

\section{Conclusions}

Network theory is demonstrably useful for analyzing organization in the U.S.~Congress. The communities arising from the legislation cosponsorship networks correlate with the ideology, geography, and committee memberships of Members of Congress. Modularity quantifies the increase in partisan polarization of the past 20 years, strengthening claims in the literature based on different data sets and methodology.  In contrast to this literature, however, modularity suggests a sharp increase in polarization \emph{prior} to the 104th Congress, indicating that it may be useful for forecasting partisan realignments.  


\section*{Acknowledgements}

We thank Thomas Callaghan, Diana Chen, Aaron Clauset, Justin Howell, Eric Kelsic, Tom Maccarone, and Mark Newman for useful discussions.  We also thank an anonymous referee for useful comments on an earlier version of this manuscript.  YZ was funded by Caltech's SURF program; AJF (VIGRE grant) and ALT (NSF HRD-0450099) were funded by the NSF; MAP was funded by the Gordon and Betty Moore Foundation through Caltech's Center for the Physics of Information (where he was a postdoc during much of this research); and PJM was supported from start-up funds provided by the Institute for Advanced Materials and the Department of Mathematics at the University of North Carolina at Chapel Hill.  We obtained the committee data from the web site of the House of Representatives Office of the Clerk \cite{clerk}.


\end{document}